\documentclass[11pt]{article}  %Do not change the point size.
\usepackage{menuproc}
\usepackage{cite}
\usepackage{epsfig}
\usepackage{amsmath,amssymb}
\def\half{\mbox{\small{$\frac{1}{2}$}}}

\def\slass#1{#1 \hspace{-1.9mm} \slash}
\def\slasss#1{#1 \hspace{-2.3mm} \slash}
\def\bbox#1{\mbox{\boldmath {$#1$}}}

\begin{document}
%
%____________________________________________________________
%
%  Title, authors, institutions, and abstract
%----------------------------------------------------------------
%  Syntax:  \titlematter{title}{authors}{institutions}{abstract}
%----------------------------------------------------------------
%     If lines are too long, use linebreaks where convenient.
%     If all authors are from the same institution, omit raised letters.
%
\titlematter{Pion-nucleon scattering in a Bethe-Salpeter approach}%
{A.D. Lahiff$\, ^a$, I.R. Afnan$\, ^b$}%
{$^a$TRIUMF, 4004 Wesbrook Mall, Vancouver, B.C., Canada V6T 2A3\\
$^b$School of Chemistry, Physics and Earth Sciences, The Flinders University
of South Australia, GPO Box 2100, Adelaide 5001, Australia}
{A covariant model of elastic pion-nucleon scattering based on the Bethe-Salpeter equation is presented. We obtain a good
description of the $S$- and $P$-wave phase shifts up to 360 MeV laboratory
energy. We also compare results from the $K$-matrix approach
and several 3-dimensional quasipotential equations to the Bethe-Salpeter
equation.}

%
%
%____________________________________________________________
%  Start article here:

%%%%%%%%%%%%%%%%%%%%%%%%%%%%%%%%%%%%%%%%%%%%%%%%%%%%%%%%%%%%%%%%%%%%%%%%%%%%%%%%%%
\section{Introduction}

Over the past decade many dynamical models of pion-nucleon ($\pi N$)
scattering based on meson-exchange have been developed. These models
invariably begin with an effective hadronic Lagrangian describing the
couplings between the various mesons and baryons. The tree-level diagrams
obtained from this Lagrangian are then unitarized using an approximation
to the Bethe-Salpeter equation (BSE)~\cite{Salpeter:1951sz}, such as
the $K$-matrix approach or a 3-dimensional (3-D) quasipotential equation. The
3-D reduction procedure is ambiguous because there are an infinite number of 
different quasipotential equations~\cite{Yaes:1971vw,Klein:1974aa}, each having 
different off-shell behaviour~\cite{Hung:2001pz}. There is no overwhelming
reason to choose one quasipotential equation over another. Sometimes
quasipotential equations have been devised so as to have the correct
one-body limit, however this argument has been shown to be not applicable
to $\pi N$ scattering~\cite{Pascalutsa:1999pv}.  Furthermore, many of 
the commonly used quasipotential equations violate 
charge-conjugation symmetry~\cite{Pascalutsa:1998vk}.
Here we avoid these problems by
constructing a covariant model of elastic 
$\pi N$ scattering~\cite{Lahiff:1999ur} based on the BSE, which we
solve without making any approximations to the relative-energy 
dependence of the kernel.

\section{The model}

The BSE for the $\pi N \rightarrow \pi N$ amplitude
$T$ is
\begin{equation}
T(q',q;P) = V(q',q;P) - \frac{i}{(2 \pi )^4} \int d^4 q''
V(q',q'';P) G_{\pi N} (q'';P) T(q'',q;P)  ,
\end{equation}
where $G_{\pi N}$ is the 2-body $\pi N$ propagator. In principle, both
the $\pi$ and $N$ propagators should be fully dressed, but for only
2-body unitarity to be maintained, $G_{\pi N}$ has the simple
form
\begin{equation}
G_{\pi N}(q;P) = \frac{1}{(\mu _{\pi} P - q)^2 - m_{\pi}^2 + i \epsilon}
\, \frac{ \mu _N \slasss{P} + \slass{q} + m_N}{(\mu _N P + q)^2
- m_N^2 + i \epsilon}  ,
\end{equation}
where $\mu _N$ and $\mu _{\pi}$ are functions of $s=P^2$ such that
$\mu _N + \mu _{\pi} = 1$. The solution of the BSE does not
depend on the choice of $\mu _N(s)$ and $\mu _{\pi}(s)$, so we use the
simplest possibility: $\mu _N = \mu _{\pi} = 1/2$.
The interaction kernel $V$ is truncated to include only the 2nd order 
$\pi N \rightarrow \pi N$ diagrams obtained from the following 
interaction Lagrangian:
\begin{eqnarray} 
 {\cal L}_{\mbox{\scriptsize int}} & = & \frac{f_{\pi N N}}{m_{\pi}} 
\, \bar N \, \gamma _5 \, \gamma ^{\mu} \, \bbox{ \tau} N \cdot
\partial _{\mu} \bbox{\pi}  + {\cal L}_{\pi N \Delta} 
+ g_{\sigma N N} \bar N N \sigma 
 + \frac{g_{\sigma \pi\pi}}{2 m_{\pi}}
\, \sigma \, \partial _{\mu} \bbox{\pi}  \cdot \partial ^{\mu} \bbox{ \pi}
\nonumber \\
& & + g_{\rho N N} \bar  N \, \half \, \bbox{ \tau} \cdot 
\left( \gamma _{\mu} \bbox{\rho} \, ^{\mu}  +
\frac{ \kappa _{\rho}}{2 m_N} \sigma _{\mu \nu} 
\partial ^{\mu} \bbox{ \rho} \, ^{\nu} \right) N 
 + g_{\rho\pi\pi} \bbox{\rho} \, ^{\mu} \cdot (\bbox{\pi} \times 
\partial _{\mu} \bbox{\pi})     . 
\end{eqnarray}
Due to the ambiguity in the description of spin-3/2 particles,
we consider two different possibilities for the $\pi N \Delta$ vertex:
\begin{eqnarray}
{\cal L} _{\pi N \Delta}^{\mbox{\scriptsize conv}} 
&=& \frac{f_{\pi N \Delta}}{m_{\pi}} \,
  \bar \Delta ^{\mu} \,\left( g_{\mu \nu} + x_{\Delta} \gamma _{\mu} 
  \gamma _{\nu} \right) \mbox{\bf T} N \cdot \partial ^{\nu} \bbox{ \pi}  
  + \mbox{h.c.}  , \\
{\cal L} _{\pi N \Delta}^{\mbox{\scriptsize Pas}} 
&=& \frac{f_{\pi N \Delta}}{m_{\pi} m_{\Delta}} \,
\epsilon ^{\mu \nu \alpha \beta} (\partial _{\mu} \bar \Delta _{\nu})
\gamma _5 \gamma _{\alpha} \mbox{\bf T} N \cdot \partial _{\beta} 
\bbox{\pi}  
+ \mbox{h.c.}  
\end{eqnarray}
Here ${\cal L} _{\pi N \Delta}^{\mbox{\scriptsize conv}}$ 
is the conventional $\pi N \Delta$ vertex which contains the so-called 
off-mass-shell parameter $x_{\Delta}$, while
${\cal L} _{\pi N \Delta}^{\mbox{\scriptsize Pas}}$
is the Pascalutsa vertex~\cite{Pascalutsa:1998pw}. 
For the $\Delta$ propagator we use the standard Rarita-Schwinger (RS) 
form~\cite{Rarita:1941mf}.
It is well known that the RS spin-3/2 propagator contains
off-mass-shell spin-1/2 components. The use of the Pascalutsa 
$\pi N \Delta$ vertex, together with the RS $\Delta$ propagator,
ensures that the $s$- and $u$-channel $\Delta$ poles present in $V$
are free of any contributions from the spin-1/2 components of the
RS $\Delta$ propagator.

\begin{table}[h]
\begin{center}
\begin{tabular}{lrrrr} \hline \hline & 
\multicolumn{2}{c}{conventional} & 
\multicolumn{2}{c}{Pascalutsa}  \\ 
 &  model I 
&  model II & model I & model II \\ \hline \hline
$g_{\pi N N}^2/4 \pi$   & 13.5  & 13.5  & 13.5 & 13.5 \\ 
$g_{\pi N N}^{(0)2}/4 \pi$  & 1.80 & 4.68 & 12.1 & 6.64 \\ 
$f_{\pi N \Delta}^2/4 \pi$  &  {\bf 0.365} & {\bf 0.365}
& {\bf 0.741} & {\bf 0.63} \\
$f_{\pi N \Delta}^{(0)2}/4  \pi$  & {\bf 0.37} & 
{\bf 0.20} & {\bf 0.193} & {\bf 0.1} \\ 
$x_{\Delta}$  & {\bf -0.11} & {\bf -0.24} & --- & --- \\
$g_{\rho\pi\pi}g_{\rho N N} /4 \pi$   & {\bf 2.88} & {\bf 2.63} & 
{\bf 2.73} & {\bf 2.25} \\
$\kappa _{\rho}$  & {\bf 2.66} & {\bf 2.03} &
{\bf 4.11} & {\bf 4.97} \\ 
$g_{\sigma \pi \pi}g_{\sigma NN}/4 \pi$  & {\bf -0.41} & 
{\bf 0.39} & {\bf -3.80} & {\bf -4.65} \\
$m_N^{(0)}$  & 1.34  & 1.14 & 1.72 & 1.18 \\
$m_{\Delta}^{(0)}$  & {\bf 2.305}  & {\bf 1.492} &
{\bf 2.60} & {\bf 1.498} \\
$m _{\sigma}$   & {\bf 0.65}  & {\bf 0.62} &
{\bf 0.69} & {\bf 1.12} \\ 
$\Lambda _N$   & {\bf 3.17} & --- & {\bf 4.90} & ---  \\
$\Lambda _{\Delta}$  & {\bf 4.56} & --- & {\bf 3.20} & --- \\
$\Lambda _{\pi}$   & {\bf 1.77} & {\bf 1.85} &
{\bf 1.76} & {\bf 2.08} \\
$\Lambda _{\rho}$  & {\bf 3.67} & --- & {\bf 3.06} & --- \\
$\Lambda _{\sigma}$  & {\bf 1.30} & --- & {\bf 4.26} & --- \\
\hline \hline
\end{tabular} 
\end{center}
\vspace*{-0.4cm}
\caption{The coupling constants and particle masses resulting from
fits to the $\pi N$ data using the two different form factor parameterizations
(denoted as model I and model II), 
and the two different $\pi N \Delta$ vertices.
The quantities in boldface were varied in the fits. All masses are in GeV.}
\label{tab:parms}
\end{table}

Regularization is achieved by the introduction of form factors. We consider
two different parameterizations: in model I we associate a cutoff function
with each vertex, where this cutoff function is taken as the product of
form factors that depend on the 4-momentum squared of each particle
present at the vertex. In model II, we associate a form factor only
with the pion propagator. In both models, each form factor is chosen to
have the form:
\begin{equation}
f(q_a^2) = \left( \frac{\Lambda _a^2 - m_a^2}
{\Lambda _a^2 - q_a^2} \right) ^{n_a}  .
\end{equation}
This form factor fulfils the requirement of only having poles along the
real axis. In this work we use $n_{\mbox{\scriptsize all}} = 1$ for
model I, and $n_{\pi} = 8$ for model II.

We solve the BSE by first expanding the nucleon propagator in the
$\pi N$ intermediate states into positive and negative energy components,
and then sandwiching the resulting equation between Dirac spinors. This
gives two coupled 4-D integral equations which are reduced
to 2-D integral equations after partial wave decomposition.
A Wick rotation~\cite{Wick:1954eu} is performed in order to obtain 
equations suitable for numerical solution. This means that 
all amplitudes are analytically continued in the
relative-energy variables from the real axis to the imaginary axis, thereby
avoiding the singularities of the kernel. Form factors with poles only
along the real axis do not interfere with the Wick rotation provided
the cutoff masses are large enough~\cite{Lahiff:1999ur}.

\section{Numerical results}

The free parameters are determined in $\chi ^2$ fits to the $S$- and
$P$-wave phase shifts up to 360 MeV pion laboratory energy from the
SM95 partial wave analysis~\cite{Arndt:1995bj}. The parameters
obtained are shown in Table~\ref{tab:parms}. Note that $g_{\pi N N}$
was fixed at $g_{\pi N N}^2 / 4 \pi = 13.5$, while $g^{(0)}_{\pi N N}$
and $m^{(0)}_N$ were determined by the nucleon renormalization
procedure~\cite{Lahiff:1999ur}. This ensures that in the 
$P_{11}$ partial wave the dressed $s$-channel nucleon pole diagram 
has a pole at the physical nucleon mass with a residue related to the
physical $\pi N N$ coupling constant.

The resulting phase shifts are shown in Figure 1. 
When the conventional $\pi N \Delta$
vertex is used, we obtain very good agreement with the partial wave analysis.
There is some disagreement in the $P_{11}$ partial wave when the
the Pascalutsa $\pi N \Delta$ vertex is used, which suggests that the
spin-1/2 components of the RS $\Delta$ propagator are necessary in order
to obtain a good fit to the $\pi N$ data.
Note that the model II results
are very close to the model I results, even though model II has four less free
parameters. Therefore in our framework the use of a different
cutoff mass for each particle results in unnecessary free parameters.
The scattering lengths and volumes are shown
in Table~\ref{tab:sl}, where it is seen that we obtain reasonable agreement
with the partial wave analyses.

Our coupling constants are consistent with the commonly accepted values
in the literature, with the exception of $f_{\pi N \Delta}$ when the
Pascalutsa $\pi N \Delta$ vertex is used. In this case, the $\pi N \Delta$
coupling constant is around twice as large as the so-called
``empirical'' value obtained from
the decay $\Delta \rightarrow \pi + N$, i.e., 
$f_{\pi N \Delta}^2 / 4 \pi = 0.36$. 
Notice in Table~\ref{tab:parms} that the choice of form factor
parameterization does not have a significant effect on the values of the
$\rho$ and $\sigma$ coupling constants. However, the choice of the
$\pi N \Delta$ vertex does make a difference to $\kappa _{\rho}$ and
$g_{\sigma\pi\pi} g_{\sigma N N}$. It has
been shown that $\Delta$ pole diagrams constructed using the conventional
and Pascalutsa $\pi N \Delta$ vertices differ 
by a contact term~\cite{Pascalutsa:2001kd}; in our
models it appears that this contact term is partially being mimicked 
by the $\rho$ and $\sigma$ exchange diagrams.

\vspace*{-0.2cm}
\begin{figure}
\hspace*{-0.5cm}
\parbox{.55\textwidth}{\epsfig{file=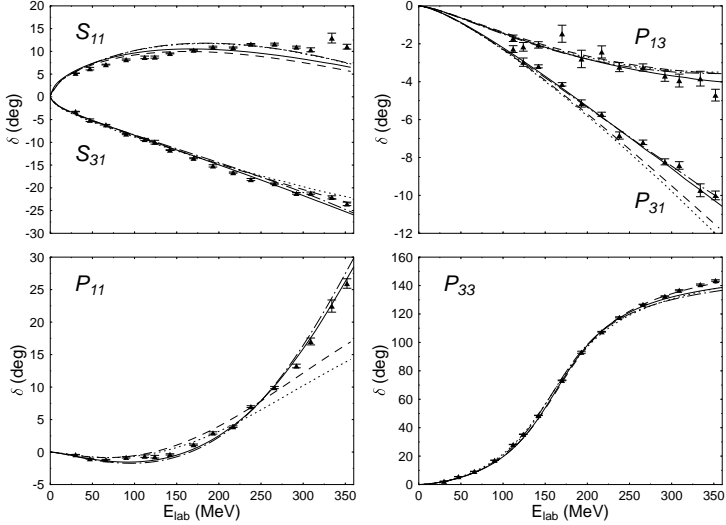,width=11.0cm}}
\hfill
\parbox{.37\textwidth}{\caption{Phase shifts obtained from the BSE using the
conventional $\pi N \Delta$ vertex [model I (solid line),
model II (dot-dashed line)], and the Pascalutsa $\pi N \Delta$ vertex
[model I (dashed line), model II (dotted line)]. The data
points are from the SM95 partial wave analysis~\protect\cite{Arndt:1995bj}.}}
\label{fig:ps}
\vspace*{-0.5cm}
\end{figure}

\begin{table}[b]
\begin{center}
\begin{tabular}{lrrrr} \hline\hline
$\ell _{2I \, 2j}$   & BSE (conv)  & BSE (Pas)  & \hspace*{1cm} SM95 
& \hspace*{1cm} KH80 \\ \hline
$S_{11}$ &  0.177  &  0.172  &  0.175 &  0.173 \\
$S_{31}$ & -0.101  & -0.105  & -0.087 & -0.101 \\
$P_{11}$ & -0.083  & -0.058  & -0.068 & -0.081 \\
$P_{13}$ & -0.032  & -0.031  & -0.022 & -0.030 \\
$P_{31}$ & -0.041  & -0.041  & -0.039 & -0.045 \\
$P_{33}$ &  0.178  &  0.187  &  0.209 &  0.214 \\ \hline \hline
\end{tabular} 
\caption{Scattering lengths and volumes obtained from the
BSE in units of $m_{\pi} ^{-(2 \ell+1)}$,
compared to results from the SM95~\protect\cite{Arndt:1995bj} and 
KH80~\cite{Koch:1980ay} $\pi N$ partial wave analyses. 
The model I form factor parameterization was used.}
\label{tab:sl}
\end{center}
\end{table}

\section{Approximations to the Bethe-Salpeter equation}

The  BSE has not often been used in meson-exchange
models to describe meson-baryon or baryon-baryon scattering
 processes. Due to the relative-energy
integration, the BSE is usually regarded as being too hard to
solve, so it has been much more common 
to use various approximations to the BSE. The $K$-matrix
approach is the simplest method, where a unitary $T$ matrix is found directly
from the on-shell potential via
\begin{equation}
T = V - i \, \mbox{Im} [G_{\pi N}] \, V . 
\end{equation}
This is obtained from the BSE when
the principle-value parts of all loop diagrams are neglected.
In Figure 2 we compare the $K$-matrix approach to the BSE,
with the parameters obtained from the BSE fits used in the
$K$-matrix approach calculations. 
For $P_{13}$ and $P_{31}$ the $K$-matrix results agree fairly well
with the BSE. The results for the remaining partial waves 
illustrate the importance of dressing and multiple scattering: 
the $K$-matrix results deviate
significantly from the BSE results. Note that for the model II form factors
the $K$-matrix approach does slightly better than for model I, i.e.,
dressing and multiple scattering are more important in model I than in 
model II. This is due to the large size of the cutoff masses
in the model I fit as compared to the cutoff mass in the model II fit
(see Table~\ref{tab:parms}).

\begin{figure}[h]
\vspace*{-0.2cm}
\hspace*{-0.5cm}
\parbox{.55\textwidth}{\epsfig{file=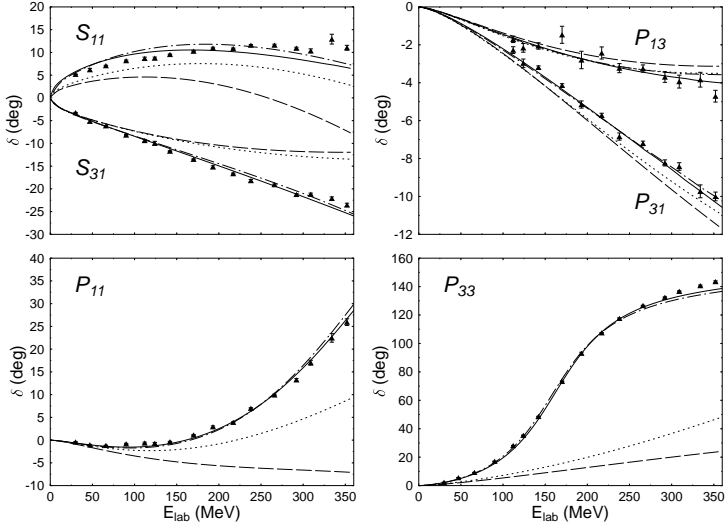,width=11.0cm}}
\hfill
\parbox{.37\textwidth}
{\caption{Comparison between the BSE [model I (solid line),
model II (dot-dashed line)], and the $K$-matrix approximation
[model I (dashed line), model II (dotted line)]. The conventional
$\pi N \Delta$ vertex was used.}}
\label{fig:kcps}
\vspace*{-0.5cm}
\end{figure}

Another approach is to approximate the relative-energy integration in 
the BSE in some way, resulting in a 3-D quasipotential equation. 
Note that many  quasipotential equations
depend on the choice of $\mu _N$ and $\mu _{\pi}$ [as defined
in Eq. (2)] due to the violation of Lorentz-invariance. Here we consider
the usual choice
\begin{equation}
\mu _N(s) = \frac{s + m_N^2 - m_{\pi}^2}{2s}  , \hspace*{1.0cm}
\mu _{\pi}(s) = \frac{s + m_{\pi}^2 - m_N^2}{2s}  .
\end{equation}

In the Cohen equation~\cite{Cohen:1970wp} it
is assumed that the $T$ matrix is independent of the relative-energy, 
and so the relative-energy integration can be performed explicitly 
over $V$ and $G_{\pi N}$, resulting in a 3-D equation. In 
Salpeter's instantaneous equation~\cite{Salpeter:1952ib}, it is 
assumed that the
interaction kernel is independent of the relative-energy, hence allowing the
relative-energy integration to be performed explicitly over just $G_{\pi N}$.
There are an infinite number of possible 3-D equations which are obtained
from the BSE by replacing $G_{\pi N}$
by an approximate 2-body propagator which generates the $\pi N$
unitarity cut, but contains a $\delta$-function on the relative-energy.
The Blankenbecler-Sugar (BbS) equation~\cite{Blankenbecler:1966gx} uses
the following propagator:
\begin{equation}
G^{\rm BbS}_{\pi N}(q;P) = 
2 \pi i \int_{s_{th}}^{\infty}
\frac{ds' f(s',s)}{s' - s - i \epsilon}  
 [ \mu _N \slasss{P} \, ' + \slass{q} + m_N] 
 \delta ^{(+)}[(\mu _N P'+q)^2-m_N^2] 
  \delta ^{(+)}[(\mu _{\pi} P'-q)^2-m_{\pi}^2]  ,
\end{equation}
where $s_{th} = (m_N + m_{\pi})^2$, and the function $f(s',s)$ is taken as 
unity. We note that the choice of $f$ is in fact arbitrary, provided
$f(s,s)=1$, hence allowing for the possibility of an infinity of
different equations. Different choices for $\mu _N$ and $\mu _{\pi}$
also result in different equations.
The Cooper-Jennings (CJ) equation~\cite{Cooper:1989ap} makes use of the
propagator
\begin{equation}
G^{\rm CJ}_{\pi N}(q;P) = 2 \pi i[ \mu _N \slasss{P} + \slass{q} + m_N]
 \int_{s_{th}}^{\infty}
\frac{ds' f(s',s)}{s' - s - i \epsilon}   
 \delta ^{(+)}[(\mu _N P'+q)^2-m_N^2] 
  \delta ^{(+)}[(\mu _{\pi} P'-q)^2-m_{\pi}^2]  ,
\end{equation}
where $f$ is chosen such that $G^{\rm CJ}_{\pi N}$ can be rewritten as
\begin{equation}
G^{\rm CJ}_{\pi N} (q;P) = 2 \pi i \, 
\frac{\delta (2 P \cdot q)}{q^2 - k^2} \, [ \mu _N\slasss{P}  + \slass{q} + m_N] ,
\end{equation}
with  $k^2 = m_N^2 - s \mu ^2_N(s)$.

Using the parameters obtained in the BSE fits to the $\pi N$ partial wave
analysis,
we calculate phase shifts using the four different quasipotential 
equations and compare the results to the BSE.
In Figure 3 we see
that all four of the considered 3-D equations agree reasonably well with the
BSE in the $S_{31}$, $P_{13}$ and $P_{31}$ partial waves. In the
$P_{11}$ partial wave, the Cohen and instantaneous equations agree well with
the BSE results, but the Blankenbecler-Sugar and Cooper-Jennings equations
generate far too much attraction. For $P_{33}$ only the Cohen equation gives
phase shifts with the same shape as the BSE results. The other three equations
produce so much attraction that the $\Delta (1232)$ resonance has become
a bound state, therefore causing the phase shifts at the $\pi N$ threshold
to be $180 ^{\circ}$, rather than $0 ^{\circ}$.

The agreement between the Cohen equation and the BSE is not surprising,
and has been found before in $\phi\phi$ scattering~\cite{Woloshyn:1973wm}.
The relative-energy integration over $V G_{\pi N}$ produces 4-body as well
as 2-body thresholds. Consequently, the $T$ matrix obtained from the
Cohen equation contains more of the analytic structure generated by
the BSE as compared to the other quasipotential equations considered here,
which only contain the 2-body threshold.

\begin{figure}[h]
\vspace*{-0.2cm}
\hspace*{-0.5cm}
\parbox{.55\textwidth}{\epsfig{file=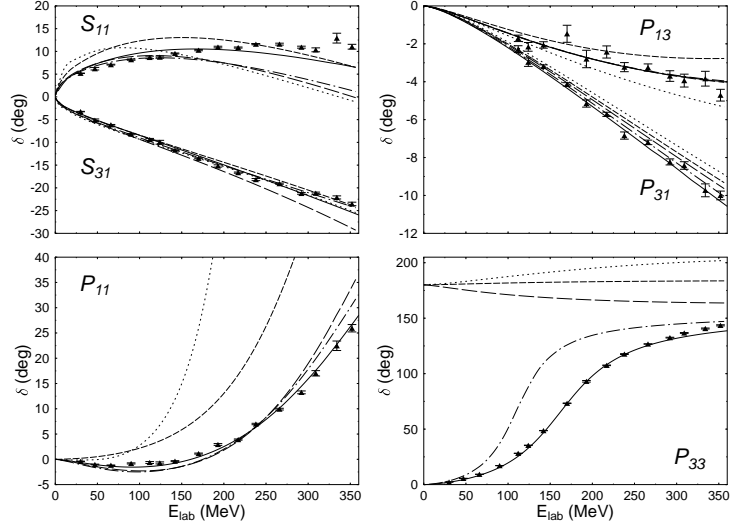,width=11.0cm}}
\hfill
\parbox{.37\textwidth}
{\caption{Comparison between the Bethe-Salpeter (solid line), Cohen (dot-dashed line),
Salpeter (long-dashed line), Blankenbecler-Sugar (short-dashed line),
and Cooper-Jennings (dotted line) equations. The conventional 
$\pi N \Delta$ vertex was used, with the model I form factor parameterization.}} 
\label{fig:3dc}
\vspace*{-0.7cm}
\end{figure}

\section{Concluding remarks}

In summary, we have presented a relativistic description 
of $\pi N$ scattering based on the Bethe-Salpeter equation, 
which we have compared
to some approximations schemes, including the $K$-matrix approach and
four different 3-D quasipotential equations.
In some partial waves large differences 
were found between the BSE and the other approaches.

The present model could be extended to higher energies by including
resonances into the interaction kernel and couplings to inelastic
channels, or extended to include the photon so as to give a
covariant model of pion photoproduction including final-state
interactions. Work in both these directions is in progress.

\acknowledgments{This work was supported by the Australian Research
Council. We also acknowledge the South Australian Centre for Parallel
Computing for access to their computing facilities.}

%%%%%%%%%%%%%%%%%%%%%%%%%%%%%%%%%%%%%%%%%%%%%%%%%%%%%%%%%%%%%%%%%%%%%%%%%%%%%%%%%%
%____________________________________________________________
%  Start references here:

\end{document}